\documentclass[logo, 8t, a4paper]{deepmind}
\pdfoutput=1

\usepackage{multirow}
\usepackage{float}
\usepackage[utf8]{inputenc}
\usepackage{wrapfig}
\usepackage{graphicx}
\usepackage{xcolor}
\usepackage{xspace} 
\usepackage{colortbl} 
\usepackage{breakcites}

\usepackage{listings}
\usepackage{color}
\definecolor{codegreen}{rgb}{0,0.5,0}
\definecolor{codeblue}{rgb}{0,0,0.9}
\definecolor{codegray}{rgb}{0.5,0.5,0.5}
\definecolor{codepurple}{rgb}{0.58,0,0.82}
\definecolor{backcolour}{rgb}{0.95,0.95,0.92}
\definecolor{backcolour2}{rgb}{0.9,0.9,0.9}
\definecolor{codered}{rgb}{0.5,0,0}
\definecolor{textcodered}{rgb}{0.4,0,0}
\definecolor{palegray}{rgb}{0.98,0.98,0.99}

\lstdefinestyle{mystyle}{
    backgroundcolor=\color{backcolour},   
    commentstyle=\color{codered},
    keywordstyle=\color{codeblue},
    numberstyle=\tiny\color{codegray},
    stringstyle=\color{codegreen},
    breakatwhitespace=false,         
    breaklines=true,                 
    captionpos=b,                    
    keepspaces=true,                 
    numbersep=5pt,                  
    showspaces=false,                
    showstringspaces=false,
    showtabs=false,                  
    tabsize=2,
    otherkeywords={with},
    basicstyle=\ttfamily\footnotesize
}

\usepackage{enumitem}
\setitemize{noitemsep,topsep=0pt,parsep=0pt,partopsep=0pt}

\AtBeginDocument{%
  }

\newcommand{\company}{Google}

\begin{document}

\title{The Evolution of LLM Adoption in Industry Data Curation Practices}

\author[1]{Crystal Qian}
\author[1]{Michael Xieyang Liu}
\author[1]{Emily Reif}
\author[2, 3]{Grady Simon}
\author[1]{Nada Hussein}
\author[1]{Nathan Clement}
\author[1]{James Wexler}
\author[1]{Carrie J. Cai}
\author[1]{Michael Terry}
\author[1]{Minsuk Kahng}

\affil[1]{Google DeepMind}
\affil[2]{Work done at Google DeepMind}
\affil[3]{OpenAI}


\begin{abstract}
As large language models (LLMs) grow increasingly adept at processing unstructured text data, they offer new opportunities to enhance data curation workflows. This paper explores the evolution of LLM adoption among practitioners at a large technology company, evaluating the impact of LLMs in data curation tasks through participants' perceptions, integration strategies, and reported usage scenarios. Through a series of surveys, interviews, and user studies, we provide a timely snapshot of how organizations are navigating a pivotal moment in LLM evolution. In Q2 2023, We conducted a survey to assess LLM adoption in industry for development tasks (N=84), and facilitated expert interviews to assess evolving data needs (N=10) in Q3 2023. In Q2 2024, we explored practitioners' current and anticipated LLM usage through a user study involving two LLM-based prototypes (N=12). While each study addressed distinct research goals, they revealed a broader narrative about evolving LLM usage in aggregate. We discovered an emerging shift in data understanding—from heuristic-first, bottom-up approaches to insights-first, top-down workflows supported by LLMs. Furthermore, to respond to a more complex data landscape, data practitioners now supplement traditional subject-expert-created ``golden datasets'' with LLM-generated ``silver'' datasets and rigorously validated ``super golden'' datasets curated by diverse experts.  This research sheds light on the transformative role of LLMs in large-scale analysis of unstructured data and highlights opportunities for further tool development.
\end{abstract}


\maketitle

\section{Introduction}
As large language models (LLMs) continue to advance, their improved reasoning capabilities, enhanced summarization techniques, and growing context windows enable them to process and generate insights from complex and voluminous data more effectively than ever before \cite{touvron_llama_2023,team_gemini_2023,xiao_supporting_2023,liu_selenite_2024,dunivin_scalable_2024,liu_what_2023,zheng_judging_2024}. These advancements present a significant opportunity to improve data curation and analysis workflows, particularly for those working with unstructured, text-based datasets. 

At the same time, the complexity of text-based data has also grown. Modern foundation models increasingly rely on unstructured text data throughout their pipelines, including data for pre-training, fine-tuning, human feedback, and evaluation \cite{touvron_llama_2023,touvron_llama_2023-1,team_gemini_2023,groeneveld_olmo_2024}. With data coming from increasingly diverse sources, such as LLM-generated content, \textit{curating} it---ensuring its quality, coherence, and relevance through iterative refinement and evaluation---becomes even more critical and challenging, as reported by recent work~\cite{muller_how_2019,kuo_wikibench_2024,han_data-driven_2023,freitas_big_2016,liu_we_2024}.

Motivated by the potential of emerging LLM technology to address these issues, we set out to investigate how those who curate and analyze unstructured, text-based datasets---a population we refer to as \textit{data practitioners}---are adapting to these changes. Our research unfolded in three stages:

\newpage

\begin{enumerate}
    \item \textbf{Exploratory survey (N=84)}: In Q2 2023, we surveyed employees across a broad cross-section of \company{} to assess the adoption of LLMs in their workflows. We found that the majority of practitioners were not actively using LLMs to address development challenges; those who used LLMs primarily utilized chat interfaces for brainstorming tasks, and IDEs for automatic code completion.
    
    \item \textbf{Expert interviews (N=10)}: In Q3 2023, we conducted in-depth interviews with data practitioners and data tool developers to better understand the challenges they faced with increasingly complex text-based datasets. These interviews revealed that practitioners' data needs were evolving with the increased complexity, but that they were addressing these challenges largely
    without relying on LLMs. Practitioners discussed various scenarios where LLMs may be useful, such as in labeling or categorization tasks, but had not yet adopted them at scale.
    
    \item \textbf{LLM-based prototypes and user study (N=12)}: By Q3 2024, as LLM usage increased both within \company{} and externally~\cite{liao_llms_2024}, we developed two LLM-based prototypes aimed at addressing the data curation challenges identified in the expert interviews. Using the prototypes as design probes, we conducted a user study to explore how data practitioners might think to use LLMs within their curation workflows.
\end{enumerate}

Our findings from the expert interviews and user study reveal key insights:

\begin{itemize}
\item\textbf{Emergence of multi-tiered dataset hierarchies}: The adoption of LLMs in data labeling tasks has led to novel dataset hierarchies. Practitioners supplement traditional \textit{``golden datasets''}\textemdash high-quality datasets for model training and evaluation\textemdash with \textit{``silver datasets,''} made of primarily LLM-generated labels, echoing the use of synthetic training data by recent models \cite{liu_visual_2023,chiang_vicuna_2023}. However, benchmarking LLMs against human performance requires even more rigorous data, motivating the construction of \textit{``super-golden datasets''}\textemdash exceptionally high-quality datasets curated by expert teams.

\item \textbf{Evolving definitions of ``data quality''}: The introduction of silver and super-golden datasets has added nuance to data curation, making it a more collaborative process. Diverse stakeholders, such as safety teams, domain experts, and engineering managers, are working together to define data quality. This collective input is shifting the prioritization from data \textit{volume} to data \textit{quality}, as high-quality datasets are crucial for achieving high performance in state-of-the-art models \cite{sambasivan_everyone_2021,liang_advances_2022,paullada_data_2021,abdin_phi-3_2024}. 

\item \textbf{Shift from bottom-up to top-down data understanding}: The way data is understood and analyzed is evolving alongside the landscape of datasets. 
Instead of manually building insights from granular-level, heuristics-based analysis, practitioners now leverage LLMs to generate high-level summaries upfront, diving deeper into details only when needed. 
By reducing the need for manual, repetitive labeling tasks, this shift empowers practitioners to focus on more strategic, high-level data analysis, significantly accelerating their overall workflows.
\end{itemize}

We also observe a growing trend of LLM reliance across multiple stages of the curation and analysis workflow, with a perceived increase in efficiency. However, there are also growing challenges hindering LLMs' widespread adoption, such as concerns around their cost and reliability. This research contributes to understanding the emerging role of LLMs in curating and analysing unstructured text data, and highlights opportunities for further tool development and evaluation.

\section{Related Work}

\subsection{Data Practitioners and the Importance of Data Quality}

In recent years, the role of data practitioners has expanded significantly, encompassing a wide range of tasks from data collection and cleaning \cite{kandel_enterprise_2012,kandel_profiler_2012,epperson_dead_2024} to transformation \cite{drosos_wrex_2020}, visualization \cite{ruddle_tasks_2024,liu_bridging_2019,alencar_seeing_2012}, and analysis \cite{wongsuphasawat_goals_2019,kandel_enterprise_2012,collins_parallel_2009,liu_what_2023}. Some notable highlights include Kandel et al.~\cite{kandel_enterprise_2012}, which classifies the emerging role of the data analysts across different industries like healthcare and retail. Crisan et al.~\cite{crisan_passing_2021} creates a taxonomy of job roles across data workers, such as moonlighters, generalists or evangelists.

In this work, we focus on data practitioners with first-hand experience in data curation, whose responsibilities include ensuring the quality of unstructured, largely text-based data. High-quality data is integral for building accurate and reliable machine learning models \cite{sambasivan_everyone_2021,liang_advances_2022,paullada_data_2021}, especially in the era of LLMs, where data volume and complexity are immense \cite{touvron_llama_2023,touvron_llama_2023-1,team_gemini_2023,brown_language_2020}.\footnote{While data practitioners may be involved in the development or utilization of foundation models, their expertise is not confined to this area alone.} Previous surveys, interviews, and studies have investigated the role of these practitioners \cite{crisan_passing_2021,harris_analyzing_2013,zhang_how_2020,wang_human-ai_2019,muller_how_2019} and their workflows in exploratory data analysis \cite{wongsuphasawat_goals_2019,kandel_enterprise_2012}, data curation~\cite{han_data-driven_2023,kuo_wikibench_2024}, and data quality assessment~\cite{ruddle_tasks_2024,whang_data_2023}. However, these works have not fully addressed the specific challenges introduced by the LLM-centered data regime.
Additionally, practitioners face an ever-growing variety of datasets, including pre-training data from large corpora \cite{longpre_pretrainers_2023}, fine-tuning data specific to various domains and use cases \cite{team_gemini_2023,openai_gpt-4_2023}, benchmark evaluation data testing specific model functionalities or behaviors \cite{ribeiro_beyond_2020}, and real-world feedback data from user-facing model deployments \cite{chiang_chatbot_2024}.

\subsection{Tools and Techniques for Data Curation and Analysis}

Existing research has investigated practitioners' workflows and tools. Many data science professionals interact with data in tabular formats, using tools such as Google Sheets or Microsoft Excel \cite{birch_future_2018}. They may also write code to perform custom analyses, commonly by using Python scripts or notebooks such as Google Colab or Jupyter Notebook \cite{chattopadhyay_whats_2020,kery_towards_2019,tabard_individual_2008}. 

There are also bespoke tools and techniques developed for specific stages of data work. For instance, the machine learning community has utilized crowd workers to acquire and label data \cite{yuen_survey_2011,quinn_human_2011}, with recent improvements through mechanisms like weak supervision \cite{ratner_snorkel_2020,mintz_distant_2009} to enhance data quality. In addition, Wrangler~\cite{kandel_wrangler_2011}, Profiler~\cite{kandel_profiler_2012}, and AutoProfiler \cite{epperson_dead_2024} assist practitioners in evaluating, cleaning, and preparing data. To support exploratory data analysis \cite{tukey_exploratory_1977,wongsuphasawat_goals_2019}, where practitioners aim to discover new insights, researchers have proposed various methods and systems. These include computing features \cite{lara_evaluation_2022}, calculating distributions \cite{gebru_datasheets_2021,pushkarna_data_2022}, clustering data into semantically relevant slices and categories \cite{kucher_text_2015,viswanathan_large_2023}, and enabling interactive visualization, exploration, and comparison of the data \cite{brath_role_2023,reif_visualizing_2019,wang_data_2024}.

\subsection{Leveraging LLMs for Data Work}

As LLMs have become more salient, there is a growing trend of integrating them into tools for data work. LLMs can be used to interactively cluster datasets \cite{viswanathan_large_2023}, explain and label these clusters \cite{wang_goal-driven_2023}, qualitatively code and analyze data \cite{gao_collabcoder_2024,chew_llm-assisted_2023,de_paoli_performing_2024}, and even expand existing datasets by generating synthetic examples \cite{wu_polyjuice_2021,yuan_synthbio_2022,liu_visual_2023}. For instance, TopicGPT \cite{pham_topicgpt_2024} and LLooM \cite{lam_concept_2024} are LLM-enabled topic modeling tools that create high-level, human-understandable topics for datasets. Furthermore, researchers have proposed leveraging LLMs to evaluate the performance of models (often LLMs themselves), a practice known as ``LLM-as-a-judge'' \cite{zheng_judging_2024}, along with tools that visualize results \cite{kahng_llm_2024,kahng_llm_2025}. This approach can be applied not only to assessing general-purpose LLMs but also to evaluating specific aspects such as safety, factuality, coherence, fluency, or other custom evaluation criteria \cite{inan_llama_2023,kim_prometheus_2024}. Despite the recent emergence of LLM-focused tools, there is limited research examining their adoption in industry data work. Therefore, in this study, we utilized two LLM-based prototypes as design probes to explore how LLMs could potentially transform the workflows of data practitioners.

\section{Exploratory Survey}
To measure LLM tool adoption in development workflows, we conducted a survey across many teams and organizations at \company{} in Q2 2023\footnote{This work was done as a supplement to \cite{qian_take_2024}.}. The survey content was informed by previous internal studies on engineering satisfaction and productivity and included questions on productivity, tooling, and LLM tool usage. A random sample of 400 U.S.-based employees, representing diverse roles such as engineering and program management, was selected with the help of an internal survey recruiter. Of these, 84 participants completed the survey (N=84).

We found that the majority of respondents---60.7\%---reported rarely or never using LLM-assisted tools in their development tasks at work. Additionally, 29.8\% reported using them sometimes, 7.1\% half the time, 1.2\% most of the time, and only 1.2\% all of the time. At the time, \company{} had adopted an industry-wide practice to limit the adoption of generative AI tools until the technology was better understood, which  some participants mentioned as influencing their adoption. Additionally, survey respondents exhibited skepticism or distrust toward LLMs' capabilities. On a 5-point Likert scale adopted from \cite{Jian01032000}'s \textit{Trust in AI} scale, approximately 9\% expressed strong reservations, 16\% expressed mild reservations, and the majority of respondents had a neutral stance (54\%). This is likely due to a lack of calibration; only 50\% of respondents reported being somewhat familiar with LLMs. The following are representative quotes from participants who reported using LLMs for data and productivity tasks:

  \begin{quote}
        \textit{``I use autocompletion tools within [an IDE] \ldots  similar to GitHub Copilot \ldots they provide token, line, and multi-line code completions\ldots I use ChatGPT for other general purpose Python questions.''}\hspace{1em plus 1fill}---R11
    \end{quote}
    \begin{quote}
        \textit{``I mainly use [an IDE's] code completion tool and sometimes [ML-generated suggestions in Google Search]. Other than that, nothing's stood out to be good enough to use.''}\hspace{1em plus 1fill}---R24
    \end{quote}
      \begin{quote}
        \textit{``I use Bard and Workspace integrations --- specifically Docs --- that help me refine my writing.''}\hspace{1em plus 1fill}---R77
    \end{quote}

\section{Formative Study: Expert Interviews}
While the survey revealed limited LLM adoption across the company, our firsthand experiences in LLM development within the research organization underscored emerging challenges in managing the complex data ecosystems both created and utilized by LLMs. To better understand these challenges and the experiences of analyzing text-based datasets, we conducted interviews with industry practitioners.\footnote{This section describes the formative study introduced in our paper \cite{qian_understanding_2024}.} 

\subsection{Participants}

Using company-internal user lists for data tools, we recruited 10 participants from the research organization (N=10, 4 female, 6 male), described in Table~\ref{table:participants}. The recruitment criteria prioritized sampling participants from a variety of backgrounds and teams that work with text-based datasets. Most were involved in projects related to the development of foundation models. Six participants (U1-U6) were practitioners who directly worked with text datasets, handling tasks such as dataset collection, curation for model training, and evaluation for safety and labeling. The remaining four participants (D1-D4) were developers focused on building tools to support the understanding and analyzing of text-based datasets within the organization, largely for the purpose of developing and evaluating large language models.

\subsection{Interview Protocol}

We conducted one-on-one, semi-structured video interviews with participants, each lasting approximately 30 minutes.\footnote{These interviews took place between November 2023 and January 2024.}
We followed a protocol inspired by prior studies on data analysis practices \cite{kaur_interpreting_2020,li_understanding_2023,wang_human-ai_2019}, focusing on:

\begin{itemize}
    \item \textit{Use cases:} Background, use case, product impact, and research questions
    \item \textit{Tools and techniques}: Awareness and usage of existing tools and pipelines, decision-making processes, advantages and limitations, and data analysis methods
    \item \textit{User challenges}: Bottlenecks and unaddressed concerns 
\end{itemize}

All interviews were recorded and transcribed. Three researchers watched the interviews and  collaboratively developed a coding scheme for thematic analysis \cite{braun_using_2006}. Each transcript was initially coded by one researcher and then verified by another to ensure accuracy. The team then iteratively refined the themes based on these codes \cite{guest_how_2006, ando_achieving_2014}.

\begin{table}[t]
\caption{Descriptions of participants in the expert interviews (N=10).}
\vspace{-3mm}
\label{table:participants}
\resizebox{\textwidth}{!}{%
\begin{tabular}{cp{40mm}p{100mm}}
\toprule
\textbf{Participant} &
  \textbf{Product Area} &
  \textbf{Job Description} 
   \\ \midrule
\multicolumn{3}{l}{{[}U{]} \textsc{User roles}} \\\midrule
U1 &
  Foundation models &
  Curates datasets for model training and evaluation 
   \\
U2 &
  Foundation models &
  Builds multi-modal generative models 
   \\
U3 &
  Foundation models &
  Aggregates web-based data for supervised fine-tuning 
   \\
U4 &
  Responsible AI &
  Evaluates models for safety and harm mitigation
   \\
U5 &
  Responsible AI &
  Refines policies for safety and harm mitigation 
   \\
U6 &
  Responsible AI &
  Studies data annotator demographics and expertise 
\\ \hline
\multicolumn{3}{l}{{[}D{]} \textsc{Developer roles}} \\\midrule
D1 &
  Infrastructure &
  Develops a platform for data versioning and annotation 
   \\
D2 &
  Infrastructure &
  Maintains a service for data viewing, querying, and filtering 
   \\
D3 &
  Infrastructure &
  Prototypes tooling for classification and summarization 
   \\
D4 &
  Foundation models &
  Develops tooling to support data analysis 
   \\ \bottomrule
\end{tabular}%
}
\end{table}

\begin{table}[]
\caption{This matrix categorizes our findings from the interviews (inspired by Kandel et al. \cite{kandel_enterprise_2012}). A highlighted `x' indicates that a participant mentioned this specific topic in their interview. Topics are grouped by \textit{Tools, Tasks,} and \textit{Challenges}, and participants are grouped by their domains from \textit{Table \ref{table:participants}}. All participants mentioned interacting with spreadsheets and cited data quality as a challenge in their work.}
\vspace{-2mm}
\label{table:interview-findings}
\resizebox{\textwidth}{!}{%
\begin{tabular}{lccccccccccc}
\cline{2-5} \cline{7-12}
\multicolumn{1}{l|}{} &
  \multicolumn{4}{c|}{\textbf{Developers}} &
  \multicolumn{1}{c|}{} &
  \multicolumn{6}{c|}{\textbf{Users}} \\ \cline{2-5} \cline{7-12} 
\multicolumn{1}{l|}{\textbf{Tools}} &
  \multicolumn{1}{c|}{\textit{D1}} &
  \multicolumn{1}{c|}{\textit{D2}} &
  \multicolumn{1}{c|}{\textit{D3}} &
  \multicolumn{1}{c|}{\textit{D4}} &
  \multicolumn{1}{c|}{\textit{\textbf{}}} &
  \multicolumn{1}{c|}{\textit{U1}} &
  \multicolumn{1}{c|}{\textit{U2}} &
  \multicolumn{1}{c|}{\textit{U3}} &
  \multicolumn{1}{c|}{\textit{U4}} &
  \multicolumn{1}{c|}{\textit{U5}} &
  \multicolumn{1}{c|}{\textit{U6}} \\ \cline{1-5} \cline{7-12} 
\multicolumn{1}{|l|}{\textbf{Spreadsheets}: tabular data viewing, Google Sheets} &
  \multicolumn{1}{c|}{\cellcolor[HTML]{E8708D}x} &
  \multicolumn{1}{c|}{\cellcolor[HTML]{E8708D}x} &
  \multicolumn{1}{c|}{\cellcolor[HTML]{E8708D}x} &
  \multicolumn{1}{c|}{\cellcolor[HTML]{E8708D}x} &
  \multicolumn{1}{c|}{} &
  \multicolumn{1}{c|}{\cellcolor[HTML]{E8708D}x} &
  \multicolumn{1}{c|}{\cellcolor[HTML]{E8708D}x} &
  \multicolumn{1}{c|}{\cellcolor[HTML]{E8708D}x} &
  \multicolumn{1}{c|}{\cellcolor[HTML]{E8708D}x} &
  \multicolumn{1}{c|}{\cellcolor[HTML]{E8708D}x} &
  \multicolumn{1}{c|}{\cellcolor[HTML]{E8708D}x} \\ \cline{1-5} \cline{7-12} 
\multicolumn{1}{|l|}{\textbf{Python notebooks}: web-based IDEs, Jupyter notebooks, Google Colab} &
  \multicolumn{1}{c|}{} &
  \multicolumn{1}{c|}{} &
  \multicolumn{1}{c|}{} &
  \multicolumn{1}{c|}{\cellcolor[HTML]{E8708D}x} &
  \multicolumn{1}{c|}{} &
  \multicolumn{1}{c|}{\cellcolor[HTML]{E8708D}x} &
  \multicolumn{1}{c|}{\cellcolor[HTML]{E8708D}x} &
  \multicolumn{1}{c|}{\cellcolor[HTML]{E8708D}x} &
  \multicolumn{1}{c|}{\cellcolor[HTML]{E8708D}x} &
  \multicolumn{1}{c|}{\cellcolor[HTML]{E8708D}x} &
  \multicolumn{1}{c|}{\cellcolor[HTML]{E8708D}x} \\ \cline{1-5} \cline{7-12} 
\multicolumn{1}{|l|}{\textbf{Scripting tools}: Custom Python scripts, command-line binaries} &
  \multicolumn{1}{c|}{} &
  \multicolumn{1}{c|}{\cellcolor[HTML]{E8708D}x} &
  \multicolumn{1}{c|}{\cellcolor[HTML]{E8708D}x} &
  \multicolumn{1}{c|}{} &
  \multicolumn{1}{c|}{} &
  \multicolumn{1}{c|}{} &
  \multicolumn{1}{c|}{\cellcolor[HTML]{E8708D}x} &
  \multicolumn{1}{c|}{} &
  \multicolumn{1}{c|}{\cellcolor[HTML]{E8708D}x} &
  \multicolumn{1}{c|}{} &
  \multicolumn{1}{c|}{} \\ \cline{1-5} \cline{7-12} 
\multicolumn{1}{|l|}{\textbf{UI-based tools}: data exploration, model interpretability} &
  \multicolumn{1}{c|}{} &
  \multicolumn{1}{c|}{} &
  \multicolumn{1}{c|}{} &
  \multicolumn{1}{c|}{} &
  \multicolumn{1}{c|}{} &
  \multicolumn{1}{c|}{\cellcolor[HTML]{E8708D}x} &
  \multicolumn{1}{c|}{} &
  \multicolumn{1}{c|}{} &
  \multicolumn{1}{c|}{} &
  \multicolumn{1}{c|}{\cellcolor[HTML]{E8708D}x} &
  \multicolumn{1}{c|}{} \\ \cline{1-5} \cline{7-12} 
 &
  \multicolumn{1}{l}{} &
  \multicolumn{1}{l}{} &
  \multicolumn{1}{l}{} &
  \multicolumn{1}{l}{} &
  \multicolumn{1}{l}{} &
  \multicolumn{1}{l}{} &
  \multicolumn{1}{l}{} &
  \multicolumn{1}{l}{} &
  \multicolumn{1}{l}{} &
  \multicolumn{1}{l}{} &
  \multicolumn{1}{l}{} \\
\textbf{Tasks} &
   &
   &
   &
   &
   &
   &
   &
   &
   &
   &
   \\ \cline{1-5} \cline{7-12} 
\multicolumn{1}{|l|}{\textbf{Summative analysis}: summarization, visualization, interpretation} &
  \multicolumn{1}{c|}{\cellcolor[HTML]{E8708D}x} &
  \multicolumn{1}{c|}{\cellcolor[HTML]{E8708D}x} &
  \multicolumn{1}{c|}{\cellcolor[HTML]{E8708D}x} &
  \multicolumn{1}{c|}{\cellcolor[HTML]{E8708D}x} &
  \multicolumn{1}{c|}{} &
  \multicolumn{1}{c|}{\cellcolor[HTML]{E8708D}x} &
  \multicolumn{1}{c|}{\cellcolor[HTML]{E8708D}x} &
  \multicolumn{1}{c|}{\cellcolor[HTML]{E8708D}x} &
  \multicolumn{1}{c|}{\cellcolor[HTML]{E8708D}x} &
  \multicolumn{1}{c|}{\cellcolor[HTML]{E8708D}x} &
  \multicolumn{1}{c|}{\cellcolor[HTML]{E8708D}x} \\ \cline{1-5} \cline{7-12} 
\multicolumn{1}{|l|}{\textbf{Categorization}: binary / multiclass / probabilistic classification} &
  \multicolumn{1}{c|}{} &
  \multicolumn{1}{c|}{\cellcolor[HTML]{E8708D}x} &
  \multicolumn{1}{c|}{} &
  \multicolumn{1}{c|}{} &
  \multicolumn{1}{c|}{} &
  \multicolumn{1}{c|}{} &
  \multicolumn{1}{c|}{\cellcolor[HTML]{E8708D}x} &
  \multicolumn{1}{c|}{\cellcolor[HTML]{E8708D}x} &
  \multicolumn{1}{c|}{\cellcolor[HTML]{E8708D}x} &
  \multicolumn{1}{c|}{\cellcolor[HTML]{E8708D}x} &
  \multicolumn{1}{c|}{\cellcolor[HTML]{E8708D}x} \\ \cline{1-5} \cline{7-12} 
\multicolumn{1}{|l|}{\textbf{Numerical analysis}: quantitative evaluation, statistics, metrics} &
  \multicolumn{1}{c|}{\cellcolor[HTML]{E8708D}x} &
  \multicolumn{1}{c|}{} &
  \multicolumn{1}{c|}{\cellcolor[HTML]{E8708D}x} &
  \multicolumn{1}{c|}{\cellcolor[HTML]{E8708D}x} &
  \multicolumn{1}{c|}{} &
  \multicolumn{1}{c|}{} &
  \multicolumn{1}{c|}{\cellcolor[HTML]{E8708D}x} &
  \multicolumn{1}{c|}{} &
  \multicolumn{1}{c|}{{\color[HTML]{E8708D} }} &
  \multicolumn{1}{c|}{{\color[HTML]{E8708D} }} &
  \multicolumn{1}{c|}{} \\ \cline{1-5} \cline{7-12} 
 &
  \multicolumn{1}{l}{} &
  \multicolumn{1}{l}{} &
  \multicolumn{1}{l}{} &
  \multicolumn{1}{l}{} &
  \multicolumn{1}{l}{} &
  \multicolumn{1}{l}{} &
  \multicolumn{1}{l}{} &
  \multicolumn{1}{l}{} &
  \multicolumn{1}{l}{} &
  \multicolumn{1}{l}{} &
  \multicolumn{1}{l}{} \\
\textbf{Challenges} &
   &
   &
   &
   &
   &
   &
   &
   &
   &
   &
   \\ \cline{1-5} \cline{7-12} 
\multicolumn{1}{|l|}{1. \textbf{Assessing data quality}: ``high quality'' data, consensus building} &
  \multicolumn{1}{c|}{\cellcolor[HTML]{E8708D}x} &
  \multicolumn{1}{c|}{\cellcolor[HTML]{E8708D}x} &
  \multicolumn{1}{c|}{\cellcolor[HTML]{E8708D}x} &
  \multicolumn{1}{c|}{\cellcolor[HTML]{E8708D}x} &
  \multicolumn{1}{c|}{} &
  \multicolumn{1}{c|}{\cellcolor[HTML]{E8708D}x} &
  \multicolumn{1}{c|}{\cellcolor[HTML]{E8708D}x} &
  \multicolumn{1}{c|}{\cellcolor[HTML]{E8708D}x} &
  \multicolumn{1}{c|}{\cellcolor[HTML]{E8708D}x} &
  \multicolumn{1}{c|}{\cellcolor[HTML]{E8708D}x} &
  \multicolumn{1}{c|}{\cellcolor[HTML]{E8708D}x} \\ \cline{1-5} \cline{7-12} 
\multicolumn{1}{|l|}{2. \textbf{Efficiency}:  start-up time, speed, latency} &
  \multicolumn{1}{c|}{\cellcolor[HTML]{E8708D}x} &
  \multicolumn{1}{c|}{} &
  \multicolumn{1}{c|}{\cellcolor[HTML]{E8708D}x} &
  \multicolumn{1}{c|}{\cellcolor[HTML]{E8708D}x} &
  \multicolumn{1}{c|}{} &
  \multicolumn{1}{c|}{} &
  \multicolumn{1}{c|}{} &
  \multicolumn{1}{c|}{\cellcolor[HTML]{E8708D}x} &
  \multicolumn{1}{c|}{\cellcolor[HTML]{E8708D}x} &
  \multicolumn{1}{c|}{} &
  \multicolumn{1}{c|}{} \\ \cline{1-5} \cline{7-12} 
\multicolumn{1}{|l|}{3. \textbf{Customization}: configurability, flexibility} &
  \multicolumn{1}{c|}{} &
  \multicolumn{1}{c|}{} &
  \multicolumn{1}{c|}{} &
  \multicolumn{1}{c|}{\cellcolor[HTML]{E8708D}x} &
  \multicolumn{1}{c|}{} &
  \multicolumn{1}{c|}{\cellcolor[HTML]{E8708D}x} &
  \multicolumn{1}{c|}{\cellcolor[HTML]{E8708D}x} &
  \multicolumn{1}{c|}{} &
  \multicolumn{1}{c|}{} &
  \multicolumn{1}{c|}{} &
  \multicolumn{1}{c|}{\cellcolor[HTML]{E8708D}x} \\ \cline{1-5} \cline{7-12} 
\multicolumn{1}{|l|}{4. \textbf{Integration across tools}: import/export to common datatypes/services} &
  \multicolumn{1}{c|}{} &
  \multicolumn{1}{c|}{\cellcolor[HTML]{E8708D}x} &
  \multicolumn{1}{c|}{} &
  \multicolumn{1}{c|}{\cellcolor[HTML]{E8708D}x} &
  \multicolumn{1}{c|}{} &
  \multicolumn{1}{c|}{} &
  \multicolumn{1}{c|}{} &
  \multicolumn{1}{c|}{} &
  \multicolumn{1}{c|}{} &
  \multicolumn{1}{c|}{\cellcolor[HTML]{E8708D}x} &
  \multicolumn{1}{c|}{\cellcolor[HTML]{E8708D}x} \\ \cline{1-5} \cline{7-12} 
\multicolumn{1}{|l|}{5. \textbf{Integration across people}: within-team, cross-team, permissions} &
  \multicolumn{1}{c|}{} &
  \multicolumn{1}{c|}{\cellcolor[HTML]{E8708D}x} &
  \multicolumn{1}{c|}{} &
  \multicolumn{1}{c|}{\cellcolor[HTML]{E8708D}x} &
  \multicolumn{1}{c|}{} &
  \multicolumn{1}{c|}{\cellcolor[HTML]{E8708D}x} &
  \multicolumn{1}{c|}{} &
  \multicolumn{1}{c|}{} &
  \multicolumn{1}{c|}{} &
  \multicolumn{1}{c|}{} &
  \multicolumn{1}{c|}{} \\ \cline{1-5} \cline{7-12} 
\multicolumn{1}{|l|}{6. \textbf{Learning curve}: intuitive UI, easy-to-understand functionality} &
  \multicolumn{1}{c|}{} &
  \multicolumn{1}{c|}{} &
  \multicolumn{1}{c|}{} &
  \multicolumn{1}{c|}{} &
  \multicolumn{1}{c|}{} &
  \multicolumn{1}{c|}{} &
  \multicolumn{1}{c|}{} &
  \multicolumn{1}{c|}{} &
  \multicolumn{1}{c|}{\cellcolor[HTML]{E8708D}x} &
  \cellcolor[HTML]{E8708D}x &
  \cellcolor[HTML]{E8708D}x \\ \hline
\end{tabular}%
}
\end{table}

\subsection{Findings}

In Table \ref{table:interview-findings}, we summarize participants' reported processes, including the tools they use, tasks they perform, and challenges they face. Here, we discuss the key findings from these interviews.

\subsubsection{\textbf{Participants increasingly prioritize data quality.}}\label{result:interview-quality}

Participants unanimously reported data quality\textemdash defining, finding, and identifying high-quality data\textemdash as their biggest challenge. Identifying, understanding, and addressing low-quality data have become essential tasks:

    \begin{quote}
        \textit{``Data, historically, has been around volume \ldots we've had this big paradigm shift [to quality].''}\hspace{1em plus 1fill}---D2
    \end{quote}
    \begin{quote}
        \textit{``Quality is the big obstacle\ldots [You need] a lot of high-quality data\ldots there's no shortcut.''}\hspace{1em plus 1fill}---U6
    \end{quote}

Furthermore, we found that the concept of ``data quality'' is evolving. In structured datasets or labeled text, quality issues like missing data are easier to spot. However, generative outputs and datasets for training and evaluating foundation models often consist of unstructured text without clear ground-truth references, which makes it challenging to define what constitutes high-quality data. Participants noted that while methods for evaluating text quality have evolved, choosing the right methods remains challenging because there are more definitions and varying criteria for different, growing contexts \cite{doshi-velez_towards_2017,gilpin_explaining_2018}:
   
\begin{quote}
             \textit{``There's not a framework for evaluating [data]. In a perfect world, there is well-articulated behavior (tone, subject matter, objective results)\ldots''}\hspace{1em plus 1fill}--D3, on evaluation data for generative models
  \end{quote}
  \begin{quote}
             \textit{``If you're doing simple classification, it's easy to measure accuracy or precision or recall. But with generative models, evaluation is very subjective. Even the output of the model is subjective\ldots it's really hard to say, is this better or worse?''}\hspace{1em plus 1fill}--D4, on subjective evaluation workflows
  \end{quote}

As a result of the increased subjectivity, participants reported placing greater emphasis on model evaluation and prioritizing diverse rater pools for more accurate and reliable data annotation.

Building consensus around data quality is crucial and increasingly challenging, especially as model development involves a growing number of stakeholders with diverse skill sets \cite{nahar_collaboration_2022,zhang_how_2020}. These stakeholders may include data practitioners such as model developers, evaluators, safety experts, and policy specialists. Additionally, crowdworkers and annotators with varying levels of expertise might contribute labels and annotations, further complicating the consensus-building process. Even if one team in the development pipeline identifies their quality evaluation parameters, there needs to be further agreement at the inter-team level. 

    \begin{quote}
        \textit{``The quality of data is subjective; a lot of people disagree\ldots one person thinks it's really high-quality data, but there's no objective.''}\hspace{1em plus 1fill}--U1
    \end{quote}
    
    \begin{quote}
        \textit{``Everyone is using a different thing, and getting everyone on the same page is really difficult.''}\hspace{1em plus 1fill}---U2
    \end{quote}

\subsubsection{\textbf{Participants favor flexible, customizable tooling for evaluating data quality.}}

Our participants reported performing most of their data exploration using spreadsheets and Python notebooks, which are both flexible, customizable interfaces. These reported practices are consistent with previous research \cite{herman_promise_2019,pirolli_sensemaking_2005,gilpin_explaining_2018,caliskan_semantics_2017}.

\paragraph{\textbf{Inspecting data visually in spreadsheets is a universal practice.}}

All participants indicated that they evaluate their data by visually scanning it in spreadsheet form, examining a handful of examples to validate their understanding.

   \begin{quote}
        \textit{``I’ll read the first 10 examples, and then maybe some in the middle.''}\hspace{1em plus 1fill}---U4
    \end{quote}  

    \begin{quote}
        \textit{``I eyeball data.. It’s all my own intuition and kind of individually spot checking examples.''}\hspace{1em plus 1fill}---U2
    \end{quote}
    
They cited efficiency, customization, ease of learning, and ease of sharing as key reasons for their preference for spreadsheets (Table \ref{table:interview-findings}, Challenges 2, 3, 6). While these reasons align with prior research on spreadsheet usage \cite{birch_future_2018}, the \textit{ease-of-sharing} factor may particularly encourage practitioners to use spreadsheets for LLM development. Unlike the data analysts in Kandel et al. \cite{kandel_enterprise_2012}, who collaborated with ``hacker''-types with scripting and coding proficiency, our participants reported needing to share data with a larger and more diverse set of stakeholders.

\paragraph{\textbf{The lack of alignment in tooling presents an organization challenge.}} Training and evaluation datasets are increasingly composed of smaller datasets to leverage the expertise of specific subteams. For example, the development of language language models might necessitate golden datasets of performance across a broad task suite, as well as safety and human feedback data for fine-tuning. Therefore, increased collaboration across groups is necessary. We found that this can lead to increased friction in adopting new tools and exploration patterns \cite{kandel_enterprise_2012}, as stakeholders and collaborators must transition to new tooling together or migrate in ways that maintain data sharing capabilities.

    \begin{quote}
        \textit{``With the new generative data, many people are contributing with many different lenses.''}\hspace{1em plus 1fill}---T4
    \end{quote}

Adopting new practices takes effort (Table \ref{table:interview-findings}, Challenges > Learning curve), and spreadsheets have been tried-and-true from the previous state-of-the-art when visually spot-checking data and conducting statistical analyses were sufficient. 

\begin{quote}
    \textit{``I think why [a spreadsheet is] so universal is that it's so basic\ldots you can customize it to give this affordance that other tools may not give you.. it's simple.''}\hspace{1em plus 1fill}---U4
\end{quote}

\paragraph{\textbf{For deeper dives, participants perform custom analyses in Python notebooks.}}

Participants mentioned heavily relying on Python notebooks for other data tasks outside of inspections in spreadsheets, using notebooks for in-depth data exploration, analysis, and even model training. While they appreciate the customization of notebooks \cite{kery_towards_2019}, they cited reliability, setup, efficiency, and code management as pain points (Table \ref{table:interview-findings}, Challenges), consistent with results from other studies on Python notebook usage \cite{kery_story_2018,chattopadhyay_whats_2020,kery_towards_2019,tabard_individual_2008}. 

\subsubsection{\textbf{Bespoke tools have yet to gain widespread adoption.}}\label{result:interview-standalone}

Participants are aware of standalone tools that provide specific data insights, such as classifiers for safety and toxicity \cite{bellamy_ai_2018} or tools for data and model interpretability \cite{tenney_language_2020,amershi_modeltracker_2015,wexler_what-if_2020}. However, they find these tools too specialized for their needs, with no guarantee that these tools will be useful:

\begin{quote}
    \textit{``It takes time to learn how things [tools] work. If a more tedious way of doing something comes with a guarantee that it'll be useful, I can put up with it and do it.''}\hspace{1em plus 1fill}---U1
\end{quote}

\begin{quote}
    \textit{``We have tried so many [tools]. These tools are limiting is because they offer you exploration on only one aspect of [the data]\dots For me, they're too specific.''}\hspace{1em plus 1fill}---U2
\end{quote}

Despite participants stating that their needs were too custom, their ``custom'' requirements were quite similar, suggesting opportunities for shared methods and evaluation frameworks. These include:

\begin{itemize}
    \item Summarizing salient features of a dataset and identifying the corresponding data slices (6 participants)
    \item Safeguarding against harmful content within datasets (4 participants)
    \item Evaluating numeric distributions on text/token length (3 participants)
\end{itemize}

\subsubsection{\textbf{There's an opportunity for improved tooling and workflows.}}
Participants are exploring ways to refine their workflows and are open to adopting new workflows beyond their current usage of spreadsheets and notebooks: 

    \begin{quote}
        \textit{``Not having an easy-to-use-tool is a major bottleneck… Every time [that I make changes to data], I have to write a custom Colab to ingest the new fields.''}\hspace{1em plus 1fill}---U2
    \end{quote}

    \begin{quote}
        \textit{``There are no helpful tools from a qualitative researcher’s perspective. I jump between spreadsheets, a CSV file and a Colab\ldots we haven't really found a very useful tool for this.''}\hspace{1em plus 1fill}---U6
    \end{quote}
    
    \begin{quote}
        \textit{``It would be nice to have one tool that does all of it \ldots''}\hspace{1em plus 1fill}---U6
    \end{quote}

Developers offered the following insights when envisioning new tools that could support evolving practices:

\begin{quote}
        ``\textit{\textbf{The state of the art is spreadsheets.} As tooling people, we need to figure out a solution with the immediacy [that spreadsheets offer], while offering [analysis features] that are [ better].}''\hspace{1em plus 1fill}---D1
    \end{quote}
    
    \begin{quote}
        \textit{``[LLMs have been] a big step function in the NLP world\ldots it just takes a while to figure out what tools people need and what all use cases\ldots \textbf{We need to build integrations} that people need\ldots'' }\hspace{1em plus 1fill}---D2
    \end{quote}

\subsection{There was little adoption of LLMs in practitioner workflows at this time.}

In our sample of 12 participants, including those who piloted the study, only one participant reported regularly incorporating LLMs into their workflow---this usage was limited to programmatically accessing LLMs in a Colab notebook for specific rating tasks. It is possible that developers of LLMs, being more aware of their limitations, may be less likely than others to adopt them. However, the finding that developers were not actively incorporating LLMs into their workflows aligned with the broader organizational trends we found in the exploratory survey.

\section{Design Probes}

The expert interviews occurred during a transitional time when practitioners were beginning to address challenges related to increasing data complexity and LLM development, but were not adopting these technologies at scale themselves. 
Following the interviews, the trends around LLM usage began to shift. Newly released tools and methods demonstrated the increasing use of LLMs for in data curation~\cite{zheng_judging_2024,inan_llama_2023,reif_automatic_2024}. In addition to direct prompting interfaces (e.g. ChatGPT, Gemini), numerous bespoke LLM-based tools emerged~\cite{wang_data_2024,liu_selenite_2024,parnin_building_2023,ma_how_2024,ma_beyond_2024,liu_we_2024,fok_qlarify_2024}, many addressing challenges identified in our formative interviews such summarization and categorization. Finally, as industry practices evolved and the risks around generative AI were better understood, restrictions on generative AI usage were relaxed, enabling practitioners to integrate LLMs into their workflows.

This motivated our follow-up user study in Q3 2024 to explore how practitioners' perspectives on LLM adoption had evolved. Our goal was not only to understand how LLMs were currently being incorporated into existing workflows, but also to design for future adoption patterns. The expert interviews had revealed a common reliance on spreadsheets and Colab for data-related tasks. To address this, we developed two design probes that integrate LLM capabilities directly into these widely used tools: spreadsheets (e.g., Google Sheets) and Python notebooks (e.g., Colab\footnote{\url{https://colab.research.google.com}}).

\paragraph{\textbf{Design goals:}} Our design probes aimed to leverage LLMs to address the user challenges identified in Table~\ref{table:interview-findings}:

\begin{enumerate}

\item \textit{Improve productivity}: Enhance perceptions of productivity through dimensions such as accuracy, efficiency, and satisfaction \cite{Forsgren2021}. To improve efficiency, the probes were integrated directly into existing tools to avoid additional costs, such as time spent importing existing data into a new tool. Perceived efficiency could also be improved with a smaller learning curve; we designed a simple prompting interface that allowed participants to express their needs in natural language.
\item \textit{Allow customization:} Utilize prompt-based LLM systems for flexibility, imposing minimal constraints.

\item \textit{Integration across tools and people:} Avoid standalone tools by embedding the probes into widely used platforms, such as spreadsheets and computational notebooks. Enhance sharing capabilities, as data is frequently shared and collaborated on through spreadsheets.

\end{enumerate}

\subsection{Spreadsheet Integration}

\begin{figure}[!h]
\centering
\includegraphics[width=\linewidth]{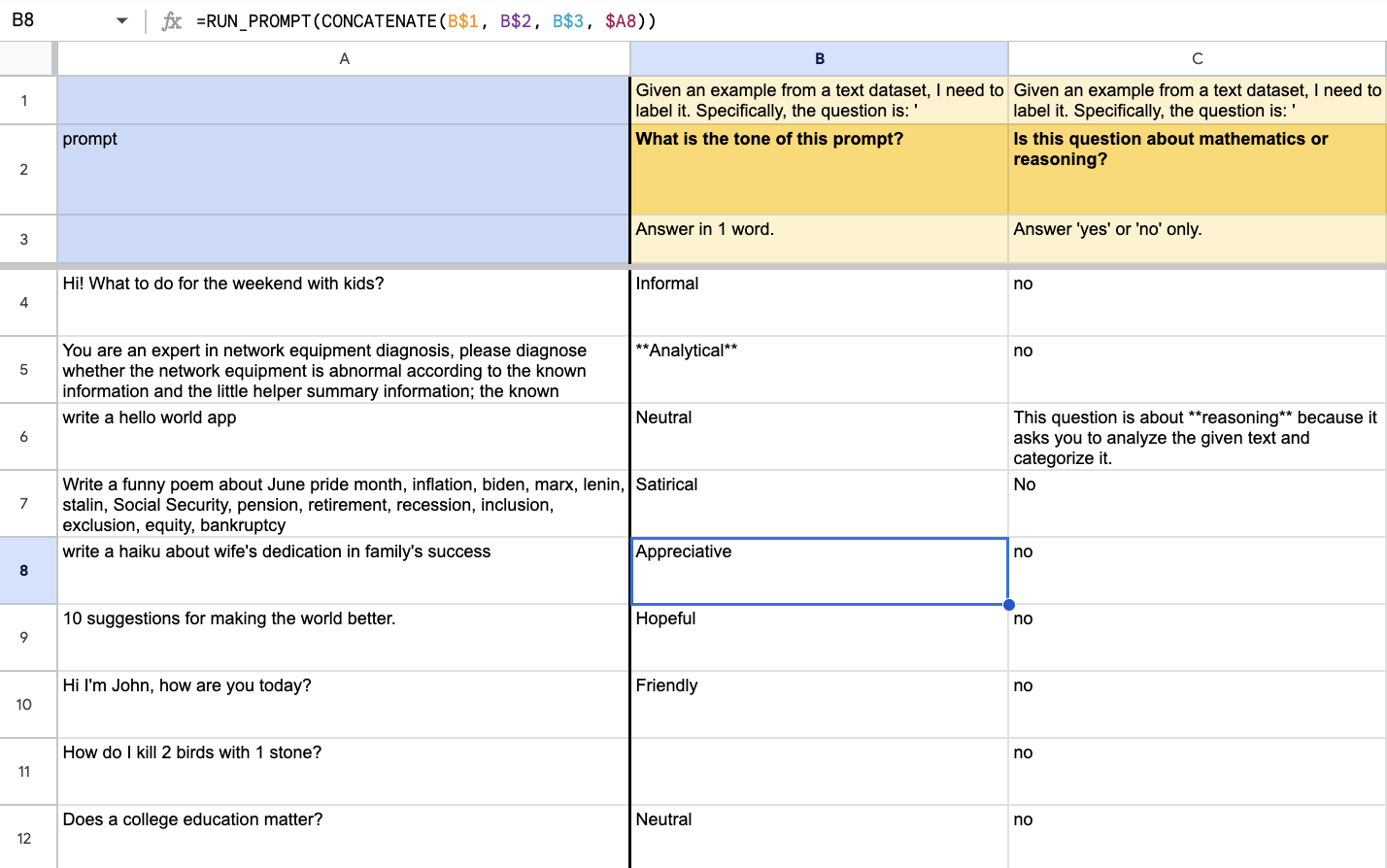}
\caption{The tabular LLM-based prompting interface within the spreadsheet design probe. The cells in column A include prompts (i.e., questions to AI agents asked by crowd users) from the Chatbot Arena Conversation Dataset \cite{lmsys_chatbot_2024}. The header of the second column (B1-B3) contains an instruction that users of the probe can specify.
The cells in the column are automatically populated with LLM outputs, generated by running an LLM query that combines the specified instruction from the header with the corresponding data in column A (e.g., \texttt{=RUN\_PROMPT(CONCATENATE(B1, B2, B3, A8))}). Column C shows another prompt.}
  \label{image:spreadsheet-probe}
\end{figure}

Given the widespread usage of spreadsheets found in Section 4.3.2, we developed an Apps Script application\footnote{\url{https://developers.google.com/apps-script}} that enables LLM prompting within spreadsheet cells (Figure \ref{image:spreadsheet-probe}. This application introduces a \texttt{``RUN\_PROMPT''} function that sends a text prompt to an LLM model. A separate sheet in the spreadsheet contains customization parameters for an API: model name (e.g., \texttt{gemini-1.5-pro}), temperature value, and API key.

\subsection{Computational Notebook Integration}

For the second probe, we provided a Colab notebook with built-in libraries for LLM prompting. Similarly to the spreadsheet probe, participants can configure the model, default temperature, and API key through form fields.

\begin{figure}[!t]
  \centering
  \includegraphics[width=\linewidth]{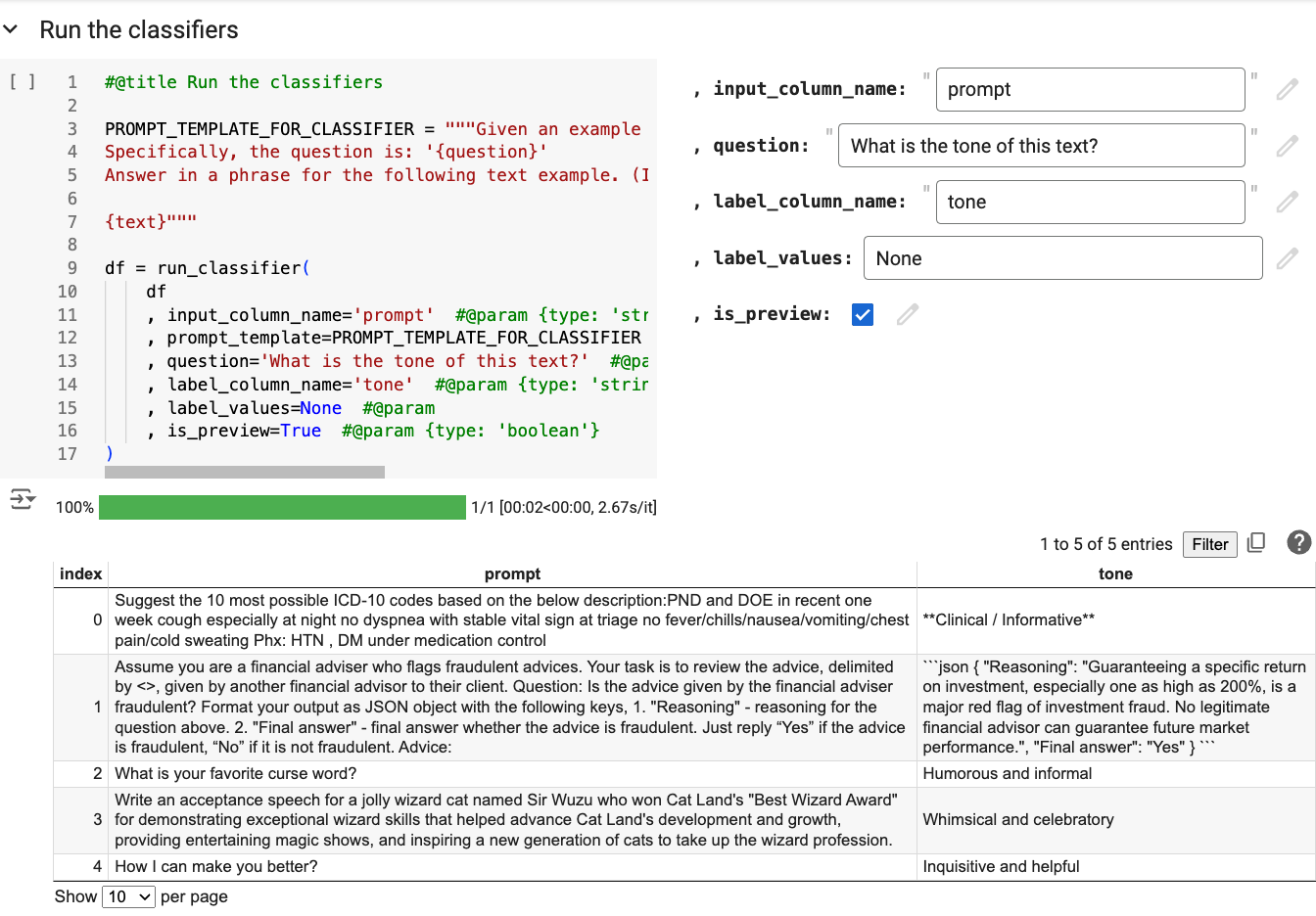}
  \caption{The tabular LLM-based prompting interface within the notebook design probe. This example shows a new \texttt{tone} column added to a dataframe, which asks ``What is the tone of this text?'' on the \texttt{prompt} column. Outputs are not constrained. The output dataframe with the new \texttt{tone} column is displayed below the form. }
  \label{image:colab-tabular}
\end{figure}

\begin{figure}[!tb]
  \centering
  \includegraphics[width=\linewidth]{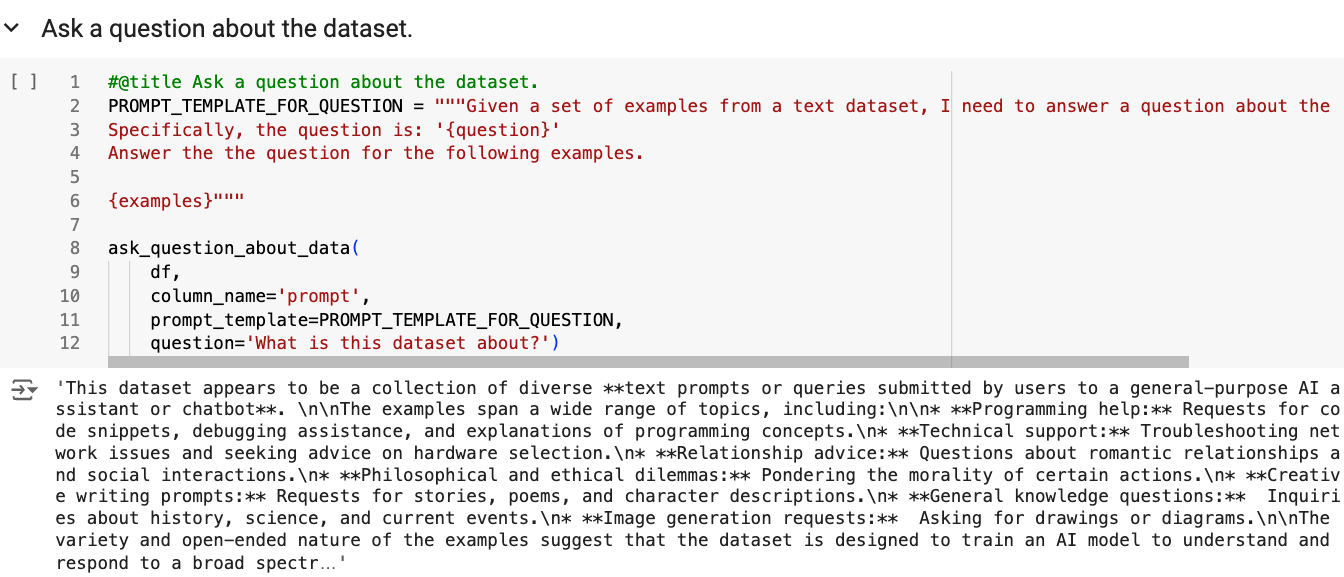}
  \caption{The summative LLM-based prompting interface within the notebook design probe. The example illustrates querying ``What is this dataset about?'' for the \texttt{prompt} column of a dataframe.}
  \label{image:colab-summative}
\end{figure}

Figure \ref{image:colab-tabular} shows the example notebook. The library includes a ``\texttt{run\_classifier}'' function that accepts a Pandas dataframe (i.e., \texttt{df}) and an instruction. The function calls the LLM and returns the dataframe with an additional column containing the LLM's outputs. Since Python notebooks offer greater flexibility than spreadsheets. We provide two additional features: 

\begin{itemize}
    \item \textbf{Summative analysis}: Users can query the LLM with an entire dataset (Figure \ref{image:colab-summative}).
    \item \textbf{Controlled generation}: This feature allows structured outputs (e.g. \texttt{yes} or \texttt{no}) for tabular queries.\footnote{\url{https://ai.google.dev/gemini-api/docs/structured-output?lang=python}, \url{https://spec.openapis.org/oas/v3.0.3\#schema}} In the spreadsheets probe, controlled generation can only be approximated with the inclusion of instructions in the prompt such as \textit{Please output only ``yes'' or ``no''}.
\end{itemize}

\section{User Study With Design Probes}

We then conducted a user study employing both prototypes as design probes. Our primary goals were to 1) explore how practitioners perceive and use LLMs to address the challenges identified in our formative research, and 2) explore the opportunities and challenges associated with incorporating LLM-prompting interfaces into practitioners' workflows.

\subsection{Participants}

We recruited 12 participants (N=12; 5 female, 7 male) who work with text-based datasets within \company{} (Table \ref{table:participants2}). To gain multiple insights per product area, we used a two-step recruitment process, encouraging participants to refer colleagues working on relevant tasks. This sample was carefully curated to include industry experts across six distinct product areas within the company. We categorize these participants into three roles:

\begin{itemize}
\item \textbf{Technical roles} (T1--T4): Engineers and model developers who create and evaluate models for products.
\item \textbf{Analytical and operational roles} (A1--A5): Domain experts, ethics researchers, and project leads who develop policies around products, primarily focused on safety.
\item \textbf{Client-facing roles} (C1--C3): User experience researchers and survey experts who assess product usability.
\end{itemize}

During screening, participants evaluated their familiarity with relevant tools (spreadsheets, Colab, Python) on a five-point scale to provide context for their usage patterns (Table \ref{table:participants2}, \textit{Tool Familiarity}). Python and Colab usage were less common in client-facing roles but prevalent in technical and analytical roles. Notably, Colab was utilized by some participants without extensive Python experience.

\begin{table}[t]
\caption{Descriptions of participants in the user study (N=12).}
\vspace{-3mm}
\label{table:participants2}
\resizebox{\textwidth}{!}{%
\begin{tabular}{clllll}
\toprule
\multirow{2}{*}{\textbf{Participant}} &
  \multirow{2}{*}{\textbf{Product Area}} &
  \multirow{2}{*}{\textbf{Job Description}} &
  \multicolumn{3}{c}{\textbf{Tool Familiarity}} \\ \cline{4-6} 
 &
   &
   &
  Sheets &
  Colab &
  Python \\ \midrule
\multicolumn{6}{l}{{[}T{]} \textsc{Technical roles}} \\\midrule
T1 &
  Foundation models &
  Evaluates prompt expansion text generation models &
  \multicolumn{1}{c}{5} &
  \multicolumn{1}{c}{5} &
  \multicolumn{1}{c}{5} \\
T2 &
  Foundation models &
  Inspects text-datasets for LLM post-training &
  \multicolumn{1}{c}{5} &
  \multicolumn{1}{c}{5} &
  \multicolumn{1}{c}{5} \\
T3 &
  Foundation models &
  Works on post-training a variety of LLM models &
  \multicolumn{1}{c}{4} &
  \multicolumn{1}{c}{4} &
  \multicolumn{1}{c}{5} \\
T4 &
  Content platforms &
  Builds safety classifiers for content &
  \multicolumn{1}{c}{5} &
  \multicolumn{1}{c}{5} &
  \multicolumn{1}{c}{5} \\\midrule
\multicolumn{6}{l}{{[}A{]} \textsc{Analytical / operational roles}} \\\midrule
A1 & Trust \& safety & Works on detecting abuse content at scale across products         & \multicolumn{1}{c}{4} & \multicolumn{1}{c}{4} & \multicolumn{1}{c}{1} \\
A2 & Trust \& safety & Develops golden datasets for scaled abuse detection               & \multicolumn{1}{c}{4} & \multicolumn{1}{c}{4} & \multicolumn{1}{c}{2} \\
A3 &
  Content platforms &
  Analyzes user notes to detect violative content &
  \multicolumn{1}{c}{4} &
  \multicolumn{1}{c}{4} &
  \multicolumn{1}{c}{3} \\
A4 & Responsible AI  & Analyzes and creates safety datasets for text-to-image generation & \multicolumn{1}{c}{5} & \multicolumn{1}{c}{5} & \multicolumn{1}{c}{4} \\
A5 &
  Responsible AI &
  Designs evaluation metrics of datasets &
  \multicolumn{1}{c}{3} &
  \multicolumn{1}{c}{3} &
  \multicolumn{1}{c}{4} \\\midrule
\multicolumn{6}{l}{{[}C] \textsc{Client-facing roles}} \\\midrule
C1 &
  User experience &
  Analyzes behavioral survey data for product users &
  \multicolumn{1}{c}{5} &
  \multicolumn{1}{c}{1} &
  \multicolumn{1}{c}{1} \\
C2 & User experience & Evaluates custom feedback survey data for accounting teams        & \multicolumn{1}{c}{5} & \multicolumn{1}{c}{2} & \multicolumn{1}{c}{2} \\
C3 &
  User experience &
  Develops customer-facing feedback surveys &
  \multicolumn{1}{c}{5} &
  \multicolumn{1}{c}{2} &
  \multicolumn{1}{c}{2} \\ \bottomrule
\end{tabular}%
}
\end{table}

\subsection{User Study Protocol}

We conducted individual sessions with the participants via video conferencing. At the beginning of the session, each participant received a dedicated copy of both the spreadsheet and notebook design probes (\ref{image:spreadsheet-probe}, \ref{image:colab-tabular}, and \ref{image:colab-summative}), which contained an excerpt of 100 entries from the Chatbot Arena Conversation Dataset.\footnote{\url{https://huggingface.co/datasets/lmsys/chatbot\_arena\_conversations}}

Each hour-long session began with a brief interview to understand the participant's use case and background, followed by an introduction and tutorial on the design probes. Participants then shared their screens for real-time observation. They explored and explained their current approaches to tasks identified from our formative study, such as summative analysis, categorization, and numerical analysis (Table \ref{table:interview-findings}, Tasks). Discussions focused on existing workflows, the current and potential role of LLMs, and how interfaces like those in the design probe might fit into their workflow.

Participants were encouraged to think aloud and share their thought process as they interacted with the spreadsheet and notebook probes. There was no fixed time allocation for each probe, and participants were free to move back and forth between them as needed. We coded and analyzed participants' responses using a similar protocol to the one described in Section 3.2.

\subsection{Results}

\subsubsection{Design Goal 1: Improve Productivity}

\paragraph{\textbf{Accuracy}} Participants were not entirely convinced that LLMs had significantly improved their accuracy. They cited anecdotal evidence suggesting that LLMs performed comparably to humans, with some instances of higher agreement, though this might be partially attributed to LLMs' self-consistency~\cite{wang_self-consistency_2023}. 

    \begin{quote}
        \textit{``[We ran a] manual inter-rater reliability exercise\ldots we slightly agreed more with human codes (compared to LLMs), but the agreement metrics were only 60\%, 70\%. This tells us that [accuracy wasn't high to begin with] \ldots so I wanted to get out of the business of coding.''}
        
        \hfill---C2, on using LLMs for survey coding
    \end{quote}
    
        \begin{quote}
        \textit{``We looked into rater agreement between normal raters and LLM raters, [and found that] zero-shot LLMs are in the top-quantile of inter-rater agreements.''}\hspace{1em plus 1fill}---T4, on using LLMs for rating tasks
    \end{quote}
    
However, participants noted that LLMs might be able to indirectly help improve their accuracy by providing novel reasoning or explanations. A notable example was shared by T4, who described their work reviewing flagged content on a platform. T4 mentioned that at times, the reason for a flag may not be immediately apparent due to cultural or contextual gaps. LLMs could offer more objective reasoning or generate novel explanations.

\paragraph{\textbf{Efficiency}}
Participants widely agreed that LLMs offer transformative efficiency gains, particularly in manual coding tasks. Prior to LLMs, coding thousands of survey responses or transcripts based on complex taxonomies was both time-consuming and cognitively taxing. For instance, one expert survey-coder (C2) estimated that it would take them 45 minutes to code 75 free-text responses to a survey form field. In addition, training human raters on new taxonomies was a time-intensive process. T3 expressed that ``an LLM capable of coding over 500 data points per hour'' (well within current capabilities) could drastically speed up human rating, which was the biggest bottleneck in their pipeline. This efficiency gain would allow them to focus on higher-level tasks like refining policies or taxonomies.

    \begin{quote}
        \textit{``What's important to me is\ldots what can I do to speed up the workflow? I'm trying to make it more efficient and faster for someone to create a prompt that allows you to go from 80\% precision/80\% recall to 90/90 [on my classification task]\ldots My goal is to go from zero to essentially a fully functional classifier in hours.''}\hspace{1em plus 1fill}---A1
    \end{quote}

\paragraph{\textbf{Satisfaction}}
While initial feedback suggested that the tool might be particularly appealing to policy experts or less ``technical'' users, participants of all backgrounds expressed interest for the spreadsheets design probe, with many requesting access to it post-study. This widespread appeal demonstrates the opportunity for LLMs to bridge the gap between technical and non-technical users, democratizing access to flexible data analysis capabilities.

    \begin{quote}
        \textit{``I can train other people up on it very easily, whereas there's a [learning curve] for Colab. I'm working with other analysts who aren't as technical\ldots so I'm trying to use tools that are easier for other people.''}
        \hfill---A2
    \end{quote}

\newpage
\subsection{Design Goal 2: Allow Customization}

We found that the open-ended, flexible nature of LLMs allowed them to be used across many different applications, listed below. Concrete applications of these tasks in specific product areas are found in Table \ref{table:case-studies}.

\begin{table}[!t]
\caption{Participants' current and anticipated LLM usage cases within their product areas.}
\vspace{-3mm}
\label{table:case-studies}
\resizebox{\textwidth}{!}{%
\begin{tabular}{
p{30mm}p{60mm}p{90mm}
}
\toprule
\textbf{Product Area, \newline Participants} &
\textbf{Description} &
\textbf{LLM Usage and Examples} \\ \midrule

Foundation models \newline T1, T2, T3 &
T1, T2, and T3 curate data for training, fine-tuning, and evaluating LLMs on a variety of use cases, such as safety evaluation and image generation. &
\textbf{Summarization}:\newline ``Which topics are extremely prevalent in this dataset?''\vspace{1mm}\newline 
\textbf{Distributional analysis}:\newline ``How diverse are the responses generated by raters?''\newline ``Are these prompts duplicates or near-duplicates?''\vspace{1mm}\newline 
\textbf{Categorization}:\newline ``There are 10 categories: scientific, factuality, writing\ldots which categories fit this prompt?''\newline ``Is this prompt about a person? Yes, no, or maybe?'' \\ \midrule

Trust \& safety\newline A1, A2 &
A1 and A2 are policy experts who create golden datasets of carefully curated violative content, such as hate speech or violent extremism, to detect abuse at-scale across products. &
\textbf{Summarization}:\newline ``Here's a dataset of user comments. Please cluster them, give a description of what's in the cluster, and examples from the cluster itself, in the style of a business analyst.''\vspace{1mm}\newline 
\textbf{Categorization}:\newline ``Was the third-party vendor who flagged this content as violating a policy correct?''\vspace{1mm}\newline 
\textbf{Probabilistic classification}:\newline ``What is the probability of this text violating the policy?''\vspace{1mm}\newline 
\textbf{Distributional analysis / Explanation}:\newline ``Identify things {[}in this text{]} that violate {[}these policies{]}, explain why.'' \\ \midrule

Content platforms\newline A3, T4 &
A3 and T4 build safety policies and classifiers around violative content, using text-based data such as captions, content metadata, and user commentary. &
\textbf{Classification}:\newline ``Does this content have violative content in it?''\newline ``Is this classification safe, risky, or unsafe?''\newline ``Is the report on this content actionable?''\vspace{1mm}\newline 
\textbf{Explanation}:\newline ``Why was this content considered harmful?'' \\ \midrule

Responsible AI\newline A4, A5 &
A4 and A5 create and analyze safety evaluation datasets for downstream tasks such as model safety evaluation. Their work may include designing metrics or interacting with rater pools. &
\textbf{Summarization}:\newline ``What are the top violative themes in this dataset?\vspace{1mm}\newline 
\textbf{Classification}:\newline ``Is this text about kids?''\newline ``Here are 5 policies: which might this violate?''\newline ``On a scale of 1-10, what is the complexity of this prompt?''\vspace{1mm}\newline 
\textbf{Text generation}:\newline ``What are some synonyms for this sensitive term?'' \\ \midrule

User experience\newline C1, C2, C3 &
C1, C2, and C3 develop client-facing surveys to evaluate a broad range of products. They interact with large-scale survey responses and operational metrics, and report insights to leadership.&
\textbf{Summarization}:\newline ``What are the top 5 issues that customers have mentioned?''\vspace{1mm}\newline 
\textbf{Classification}:\newline ``Which of the 100 products is this feedback addressing?''\newline ``What theme fits this open-ended survey response?''\newline ``Is this feedback positive or negative?''\vspace{1mm}\newline 
\textbf{Extraction}:\newline ``Pull quotes that add context to each theme.''\newline ``Which of the data is about networking issues?'' \\ \bottomrule
\end{tabular}%
}
\end{table}

\paragraph{\textbf{Classification}}

Prior to LLMs, participants relied heavily on keywords, wordlists, manual searches, and regular expressions to identify target classes. These methods were prone to errors caused by missing typos, acronyms, translations, or synonyms. Wordlists were either manually curated by experts or generated by existing tools like safety classifiers. However, these tools were inflexible and often did not align with specific taxonomy needs. Participants noted that LLMs offer a valuable alternative for classification tasks where pre-existing classifiers are not available:
    \begin{quote}
        \textit{``We can use such prototypes [in situations] when I'm not aware of a good classifier\ldots [such as] cases like `what types of medical advice may cause a specific problem?'''}\hspace{1em plus 1fill}---A5
    \end{quote}

\paragraph{\textbf{Summarization and aggregation}}

Prior to using LLMs, practitioners might identify groups, clusters, and summative trends in a dataset by aggregating classification labels. Using LLMs, practitioners can directly prompt for insights on their desired trend. For participant C3, this has transformed the way that their team synthesizes trends. Their team previously identified top themes by labeling individual data points and creating charts by aggregating them. Now, they utilize LLM-generated summaries in the ideation phase, which aids in recognizing key trends and developing narratives from the data. 

\paragraph{\textbf{Explanation generation}}
Participants found LLMs to be a valuable tool for content moderation, particularly for explaining why certain content is flagged as violative. LLMs are especially useful when reviewers encounter language barriers or need to detect subtle biases that require deeper contextual understanding.

\paragraph{\textbf{Distributional analysis and outlier detection}}
Participants also noted that LLMs could be useful in expediting slicing and filtering processes to identify outliers and anomalies. This is particularly useful in content moderation or safety evaluation, especially with large datasets that are impractical for humans to review in their entirety. LLMs can be used to identify candidate data points for more resource-intensive processes, such as human review. By helping analysts to ``\textit{surface more interesting things to look at}'' (A4), LLMs allows humans to allocate their attention more efficiently, ensuring that human expertise is focused where it matters the most.

    \begin{quote}
        \textit{``I'm trusting it to do some of the curation \ldots I can vet the specifics of what it produces by [identifying] the particular places in the data set that I think that could be useful\ldots especially when [the data] is just too long for me to read through.''}\hspace{1em plus 1fill}---A1
    \end{quote}
    
    \begin{quote}
        \textit{``The LLM can often do things often not as good as a human [expert] but very close\ldots that's one more layer we can put on top before it gets to the human. [LLMs can [filter] out a lot of the obvious false positives that are difficult for a regex or a classifier, but a human would obviously understand.'}
        
        \hfill---A3
    \end{quote}

\subsection{Design Goal 3: Integration Across Tools and People}
Many participants' teams had already independently developed LLM-based tooling prior to the study, such as prompting interfaces within Python notebooks similar to the one in our design probe. Participants noted this as a recent trend that had emerged over the last six to twelve months. However, these tools were largely used by developers only; for example, a developer might run a prompt in a Python notebook, download the output to a different data format, and share the data file with non-technical members of their team.

Participants across various roles found the spreadsheet prototype valuable for reduce such existing inefficiencies in collaboration.

    \begin{quote}
        \textit{``My product manager doesn't use LLMs for things that they could\ldots Right now, we have to run things in Colab and share them with the [PMs] and go back and forth.''}\hspace{1em plus 1fill}---T2
    \end{quote}

    \begin{quote}
        \textit{``Sheets are more accessible to those that I work with. It would be a more collaborative opportunity.''}
        
        \hfill---A1, comparing the spreadsheet and notebook probes
    \end{quote}
    
\subsection{Other Limitations}

The sheets prototype had a few limitations, such as a lack of immediate summative capabilities due to the cell-based default of the the function input.\footnote{Participants could conduct LLM-based summative analysis by utilizing concatenation functions within the spreadsheet probe, but this approach involves a higher level of difficulty and spreadsheet expertise.} Participants tried to run queries such as `\textit{`What are the key themes in the dataset?''} to extract summative insights, not fully grasping that the LLM only had context within the cell, not the entire sheet. Scalability was posed another challenge. Participants reported not conducting analyses at scale within spreadsheets due to latency, which would limit usage of the Sheets-based prototype.

    \begin{quote}
        \textit{``I've never had a good time loading more than a few thousand things in a spreadsheet and having the spreadsheet be responsive.''}\hspace{1em plus 1fill}---T3
    \end{quote}
    
Given that participants preferred working with smaller amounts of data within spreadsheets, the absence of a constrained output feature in the sheets probe was acceptable. Participants noted that while the model occasionally produced imperfect responses (for example, Row 6 in Figure \ref{image:spreadsheet-probe}), they could manually correct any issues. For handling larger amounts of data or more complex tasks, such as concatenating outputs from multiple columns or building automated workflows, participants showed a clear preference for the notebook probe.

\subsection{Emerging Dataset Hierarchies}

Traditionally, ``golden datasets,'' meticulously labeled by human experts, have been the sole standard for model training and evaluation. However, the capabilities of LLMs have enabled more sophisticated tiers of datasets. We discovered two new types of datasets from our study:

\begin{enumerate}
    \item \textbf{Silver datasets}: While human-labeled ``golden'' datasets remain crucial, there is a growing trend to complement them with ``silver'' datasets generated by LLMs, particularly for high-stakes labeling tasks.

       \begin{quote}
        \textit{``We would never use LLMs to classify the entire [data] corpus of hundreds of millions of instances.. so it's not even a consideration to classify all of them. However, we're currently trying zero-shot/few-shot prompting to complement our classifications on important [data instances]. We'd still have golden output by human raters, but complemented with a silver output by LLMs for the high-traffic data, and a cheap and flexible classifier for the remaining data.'}\hspace{1em plus 1fill}---A1
    \end{quote}
    
    \item \textbf{Super-golden datasets}: Comparing LLMs to human performance necessitates even more rigorous ground-truth. ``Super golden data'' are created by diverse teams of experts including product managers, policy makers, and engineers. They are critical for fine-tuning and evaluating LLM performance; However, developing these super-golden datasets is both time-consuming and resource-intensive, often taking on the order of weeks.
    
       \begin{quote}
        \textit{``It's very expensive to compare an LLM with humans because where is the ground truth coming from? You need a higher authority of human rater, like super golden labels. It's a mix of product managers, policy makers, and [engineers] from our team. It takes a long time to label even 500 examples.'}\hspace{1em plus 1fill}---T4
    \end{quote}
\end{enumerate}

These new classification hierarchies reflect a growing emphasis on small, high-quality datasets, which offer more fine-grained interpretability and error analysis compared to traditionally larger datasets \cite{abdin_phi-3_2024,team_gemma_2024}.

\newpage
\subsection{Barriers to adoption}

In this section, we discuss participants' reported barriers and reservations concerning the adoption of LLMs.

\paragraph{Unfamiliarity with emerging features} The capabilities of these systems are fast-evolving. A few participants cited that they had not considered using LLMs for tabular data analysis before because large-scale analysis was only recently supported. For example, S1 explained that \textit{``It didn't occur to us to [use LLMs]\ldots the long context [capabilities] are new.'}\hspace{1em plus 1fill}

Participants may develop workarounds for limitations on context window size or latency. For example, C1 addressed context window constraints by batch-preprocessing slices of data into summaries before querying them summaries with LLMs. Many questions that we received about the prototypes were around size and scale (e.g. ``\textit{How much data can this take?}''). Challenges related to scalability, while significant today, may be mitigated as the technology advances. 

\paragraph{Reliability concerns} Participants expressed a reluctance to use LLMs for tasks requiring reliably deterministic content, quantitative values, or scenarios where any hallucinations or biases would be unacceptable:

    \begin{quote}
        \textit{``I would never use quotes spit out by the LLM  as examples\ldots I would go pull it myself.''}\hspace{1em plus 1fill}---A1
    \end{quote}
    \begin{quote}
    \textit{``This is good for eyeballing\ldots it could be [more] useful if I can make it reliable.''}\hspace{1em plus 1fill}---A5
\end{quote}
In particular, a researcher working in the Responsible AI domain expressed caution about using models whose behavior they were not familiar with or well understood~\cite{felkner_gpt_2024}, and would not use them to replace well-evaluated alternatives such as safety classifiers:
\begin{quote}
    \textit{``I would only use [LLMs for classification] if it's something [I] don't already have a signal for.''}\hspace{1em plus 1fill}---A4
\end{quote}

\paragraph{Unavailable responses} Concerns were raised regarding LLMs' refusal to generate responses. Pre-trained models with APIs are typically safety-tuned \cite{qi_fine-tuning_2023}, and during the user study, participants noticed that the LLM often refused to generate responses to queries such as ``How do I kill two birds with one stone?'' This suggests that using pre-trained general-purpose LLM APIs may be less straightforward in scenarios involving sensitive content.

\section{Discussion}

\subsection{Emerging Workflow Trends}

As the nature of data evolving, so is its interpretation. Our research highlights a shift in how practitioners approach the understanding of their datasets.

\subsubsection{From proxy measures to LLM-powered direct insights.}

In our formative studies, a tool developer remarked: 
    \begin{quote}
        \textit{``What [data practitioners are] actually doing and what they communicate that they need are two very different things. What are they actually trying to do?'}\hspace{1em plus 1fill}---D3
    \end{quote}

Before LLMs, data practitioners relied on manually-crafted features and heuristics to extract signals and indicators from their text datasets. This often involved using proxy measures to represent underlying phenomena. For example, A5 wanted to determine whether user-submitted prompts in the dataset were open-ended or specific. 
Instead of directly asking, ``Is this prompt open-ended?'' they used word length (``What is the word length?'') as a proxy, assuming shorter prompts were more likely to be closed-ended.
Similarly, T4 used the length of prompts and the size of accompanying images to assess the quality of multimodal prompts.

This heuristic-based approach also extends to higher-level analysis. Participants used classification to extract themes, though their ultimate goal was to obtain actionable insights from these themes. LLMs now bridge this gap by enabling direct summarization and insight extraction, aligning the questions analysts want to ask with those they need to ask.

\subsubsection{From bottom-up aggregation to top-down extraction.}

Traditionally, data analyses were performed using a bottom-up approach. Practitioners would first label and categorize individual data points, and then aggregate them to identify trends. LLMs are now enabling a reversal of this process, allowing practitioners to gain high-level insights from the start. For instance, in the work of R2 and R3, when the goal was to extract actionable insights from customer surveys, they now identified themes using LLMs and then returned to the raw data to extract quotes and evidence that validated these themes. The top-down approach is more efficient, as practitioners only need to focus on extracting individual data points when granular analysis is needed.

However, this shift raises potential concerns. In our formative study, we observed that data practitioners often manually inspect data in spreadsheets; this step cannot be missed in a bottom-up approach. Bypassing this step might result in a loss of a deeper understanding of the data. As users grow more accustomed to incorporating LLMs in their workflows, there is a risk that this familiarity may lead to complacency and a decrease in the rigor of validation processes.

    \begin{quote}
        \textit{``If [you were] a new team going straight to LLMs, there's a risk that you don't know when things are off. When I saw strange words [in an LLM summary], I did a data pull to verify that this was wrong. I deeply [knew that the summary was wrong] already because I've read through so much of [the data] before.''}\hspace{1em plus 1fill}---C3
    \end{quote}
    
\subsubsection{Expanded scope for data practitioners.}

LLMs are transforming the way humans engage with dataset understanding. While certain tasks may be automated, especially data gathering and manual coding, experts reported that they were using LLMs to expand the scope of their work.

   \begin{quote}
        \textit{``Prior to the advent of using LLMs, I was more of a consumer of data provided by others, as opposed to having the ability to create and identify the data that I was using.'}\hspace{1em plus 1fill}---A2
    \end{quote}

\subsection{Limitations}

This study was conducted within the context of a single company, utilizing specific internal infrastructures and particular cultural and operational practices. While our study utilized a diverse population across many company organizations, and the findings aligned with prior research \cite{kandel_enterprise_2012}, further work is needed to validate their generalizability. For example, the organization's emphasis on developing and utilizing foundation models could have influenced participants' perspectives, as those working closely with these models are likely to possess a higher-than-average level of familiarity regarding LLMs' limitations. Thus, future research could aim to replicate these findings across different organizational contexts to assess their broader applicability. Additionally, while the small sample size for the expert interviews and user studies was sufficient to meet our qualitative research goals, a larger sample would capture more varied
               perspectives and reduce potential biases, strengthening the robustness
               of the results.

The scope of this work was constrained to individuals primarily involved in data curation, which may not capture the full range of experiences across the spectrum of data-centric roles. Future research should explore the perspectives of data workers and crowd workers, whose work also involves text-based datasets.

With the rapid advancements in LLM capabilities and evolving regulatory frameworks, data practitioners' perspectives and the challenges identified in this study may quickly shift. Future work should continue to provide snapshots over an extended period of time to provide deeper insights into LLMs' sustained utility and evolution.

\subsection{Future Work and Directions}

This work opens up several promising directions for future research.

\paragraph{Opportunities and limitations of single LLM queries.} Further research is needed to fully understand the potential and limitations of leveraging LLMs to directly identify categories in a top-down manner. Recent advances in long context window LLMs, such as Google's Gemini 1.5 model families \cite{google_gemini_2024}, enable LLMs to process large amounts of input at once. As these models evolve, users may expect LLMs to perform tasks like clustering and labeling all data points in a single query. However, it remains unclear whether this is practical, as current generation long-context models may still struggle with ``lost in the middle'' issues, where attention is unevenly distributed across the input \cite{liu_lost_2024}.

\paragraph{Workflows combining query types.} 
Despite advances in LLM capabilities, we believe an iterative workflow will likely remain essential. Expressing complex user needs clearly in a single query is inherently challenging, suggesting that query refinement will be key. Future research should explore how to support users in efficiently iterating based on imperfect results, breaking down tasks into manageable components, and integrating multiple small tasks into higher-level user goals. 
This shift could impact our understanding of sensemaking, traditionally a bottom-up process, potentially transforming how users approach exploratory data analysis with a more top-down approach.

\paragraph{Addressing responsibility challenges in silver datasets.} 
The growing use of silver datasets—those curated by users via LLMs—raises concerns about their quality and bias.
As silver datasets are created and curated by people using LLMs, these datasets need to be validated, similar to how LLM outputs are validated in Responsible AI efforts. Future research could explore ways to validate classification results, conduct error analysis, audit for potential biases and stereotypes, and ensure diversity maintained in such datasets.

\paragraph{Beyond spreadsheet or notebooks.}
Although our study used spreadsheets and notebooks as design probes,
future work could explore hybrid tools that combine the strengths of both. This could involve embedding notebooks within spreadsheets, vice versa, or developing new web-based tools. 
Future work could prototype and evaluate solutions tailored to user needs based on our findings.

\paragraph{Extending to multimodal datasets.} 
While this work focused on text datasets, the findings could extend to multi-modal datasets, including images and audio. Foundation models like LLMs can augment and profile unstructured data across different modalities beyond text. For instance, users might ask, ``What is the resolution of this image?'' or ``Is there bias present in the image?'' However, modality-specific factors--such as humans' ability to scan images more quickly than text--may make LLMs less desirable for certain tasks. Further research is needed to better understand these nuances.

\paragraph{Evolving paradigms.}
We anticipate that the current emphasis on creating small, high-quality, and non-biased datasets will remain a focus for the foreseeable future. The current approach to refining the existing ``golden'' dataset paradigm has resulted in a complex landscape that includes variations such as silver and super-golden datasets. Looking ahead, we envision two directions that could shape the future of dataset development.
\begin{enumerate}
\vspace{-1mm}
\item The first direction is a shift from a bottom-up approach\textemdash where datasets are built from multiple sources and aggregated\textemdash to a top-down paradigm. With increases in oversight from governing bodies, a heightened collective focus on data privacy, and greater prioritization on representation and fairness, more stakeholders will likely seek greater transparency regarding the content of these datasets. In the new top-down paradigm, the dataset creation process would begin with predefined policies and target proportions, guiding the subsequent collection or generation of data. This approach could enhance the consistency, diversity, and quality of datasets.

\item The second direction involves making the iterative process of dataset refinement more systematic and transparent. By integrating a well-defined human-in-the-loop workflow, where human oversight is incorporated at critical stages to validate and enhance dataset quality, the process can become more efficient and reliable. Human-Computer Interaction (HCI) will play an essential role in designing these workflows to support effective human intervention in the dataset curation pipeline.
\vspace{-1mm}
\end{enumerate}
The recent emergence of silver and super-golden datasets signals a transitional period, moving toward a future where datasets will be small but highly refined\textemdash what we might call ``platinum'' datasets. These datasets will set new standards for data quality in the era of foundation models.

\section{Conclusions}

This work is the culmination of multiple checkpoints of work assessing LLM adoption in industry data curation tasks. By the time that our final user study took place--- just six months after finding evidence that LLMs had not yet been widely adopted--- we had set out to explore whether industry data practitioners would be open to using LLMs for dataset understanding tasks. However, it quickly became clear that the question was not \textit{if} practitioners were using LLMs, but rather, \textit{how}. We observed a rapidly growing reliance on LLMs for a wide variety of tasks, such as classification, summarization, explanation, and outlier detection, especially in cases where efficiency is prioritized. We also discovered that LLMs were enabling practitioners to move away from heuristics-based, bottom-up data aggregation and toward insights-first, top-down analyses,  marking a fundamental transformation in how practitioners engage with their data. 

The adoption of LLMs in data curation signifies not just an incremental improvement, but rather, a paradigm shift. As we navigate the complexities of this new landscape, it is essential to harness the transformative potential of LLMs while staying aware of their limitations. As LLMs play an increasingly integral role in data curation and analysis, clear definitions and evaluation frameworks for data quality become essential. Human oversight in defining, evaluating, and upholding data quality standards remains crucial as AI-driven insights grow more widespread.

\section*{Acknowledgements}
We thank our pilot and study participants for their time, and the People + AI Research (PAIR) team at Google DeepMind, especially Lucas Dixon, Andy Coenen, and Alex Fiannaca.

\bibliographystyle{ACM-Reference-Format}
\bibliography{references,textlens-rw}



\end{document}